# Effects of Disorder on Charge Orderings and Superconductivity in the System of Coexisting Itinerant Electrons and Local Pairs

*Stanisław Robaszkiewicz and Grzegorz Pawłowski*

*Institute of Physics, A. Mickiewicz University*

*ul. Umultowska 85, 61-614 Poznań, Poland*

## Abstract

We study the phase diagrams and thermodynamic properties of a system of coexisting itinerant electrons and local pairs (LP), in the presence of diagonal disorder. The model considered takes into account both the charge exchange couplings, responsible for superconducting orderings, and the density-density interactions, which can stabilize charge ordered states in the system. Depending on the strength of the random on-site potential, the interactions and the particle concentrations the model is found to exhibit several various phases, including the homogeneous ones: superconducting (SC), charge density wave (CDW) and nonordered (NO), as well as the phase separated states (CDW-SC and CDW-NO).

The model of coexisting itinerant electrons and bound electron pairs i.e. the boson-fermion model, has been recently proposed for a description of the superconducting copper oxides and other nonconventional (exotic) superconductors, the chalcogenide glasses, as well as some bipolaronic systems [1-3]. Up to now the studies of the model have been concentrated on the superconducting (SC) and nonordered (NO) phases [1-6].

In this paper we extend the earlier investigation by including the charge ordered (CDW) states, which can be stabilized in the system by the density-density interaction (and/or the electron-phonon coupling). We point out the properties of these states, the competition between CDW and SC orderings and discuss the effects of the diagonal disorder in LP subsystem on the relative stability of various phases possible in the system.

We consider the following model Hamiltonian:

$$H = 2\sum_i (\Delta_o + E_i - \mu)n_i^b - \sum_{<ij>} J_{ij} b_i^+ b_j + 2\sum_{<ij>} K_{ij} n_i^b n_j^b +$$

$$+ \sum_{<ij\sigma>} t_{ij} c_{i\sigma}^+ c_{j\sigma} - \mu \sum_i n_i^c + I_o \sum_i (c_{i\uparrow}^+ c_{i\downarrow}^+ b_i + h.c.) + V_o \sum_i n_i^b n_i^c \quad , \quad (1)$$

$\Delta_o$ measures the relative position of the local pair (LP) level with respect to the bottom of c-electron band, $E_i$ is the random LP site energy, $\mu$ stands for the chemical potential, $I_o$ and $V_o$ are the intersubsystem charge exchange and density-density interaction respectively, $J_{ij}$ and $K_{ij}$ denote the LP transfer integral and the LP density–density interaction, respectively. The charge operators for LP (hard-core bosons) $b_i^+$, $b_i$ obey the Pauli spin 1/2 commutation rules [1 – 2], $n_i^c = \sum_\sigma c_{i\sigma}^+ c_{i\sigma}$, $n_i^b = b_i^+ b_i$, $\mu$ is the chemical potential which ensures that a total number of particles is constant, i.e. $n = n_c + 2n_b = (\sum_i <n_i^c> + 2\sum_i <n_i^b>)/N$. We will not consider here the randomness in fermion site energies, since weak disorder of this type is responsible mainly for renormalization of the single particle density of states (DOS).

The model was analysed within the mean field variational approach (MFA-HFA) similar as that used earlier for the model (1) without disorder [1]. First, $F_o(\{E_i\})$, i.e. the variational free energy for a given fixed configuration of the random site energy $\{E_i\}$ is obtained. Then it is configurationally averaged over the random variable $\{E_i\}$ according to a preset probability distribution $P(\{E_i\})$ as $<...>_{av} = \int_{-\infty}^{\infty} \prod dE_i P(\{E_i\}) \ldots$.

The probability distribution $P(\{E_i\})$ of $\{E_i\}$ is assumed to be $P(\{E_i\}) = \prod_i p(E_i)$, with $p(E_i) = p(-E_i)$ [7] and in our work we consider two types of energy distribution: *the two-delta distribution:* $p(E_i) = (1/2)[\delta(E_i - E_o) + \delta(E_i + E_o)]$, and *the rectangular distribution:* $p(E_i) = 1/2E_o$ for $|E_i| \leq E_o$, $p(E_i) = 0$, *otherwise*.

Assuming the existence of two interpenetrating sublattices (A and B) and restricting the analysis to the two-sublattice solutions [1, 7] we have derived the quenched free energies $<F_o(\{E_i\})>_{av}$ and selfconsistent equations determining order parameters and $\mu$ for various ordered phases possible in the system, including also the phase separated states.

We have taken into consideration the following homogeneous phases:

(1) singlet superconducting (SC) with $\rho_o^x \neq 0$ and $x_o \neq 0$, where

$$\rho_q^x = \frac{1}{2N} \sum_i << b_i^+ + b_i >>_{av} e^{-i\vec{q}\vec{R}_i} \quad \text{and} \quad x_q = \frac{1}{N} \sum_k < c_{k+q\uparrow}^+ c_{-k\downarrow}^+ >,$$

(2) charge ordered (CDW) with

$$\rho_Q^z = \frac{1}{N} \sum_i << b_i^+ b_i >>_{av} e^{-iQR_i} \neq 0 \quad \text{and} \quad n_Q = \frac{1}{N} \sum_{k\sigma} < c_{k+Q\sigma}^+ c_{k\sigma} > \neq 0, \quad \vec{Q} = (\frac{\pi}{a}, \frac{\pi}{a}, \frac{\pi}{a}),$$

(3) nonordered (NO), (4) mixed CDW-SC with $\rho_Q^z \neq 0$, $n_Q \neq 0$ and $\rho_q^x \neq 0$, $x_q \neq 0$ ($q = 0$, $q = \vec{Q}$), as well as the phase separated states: (5) CDW-SS phase separated (PS1), (6) CDW-NO phase separated (PS2).

We have performed an extended analytical and numerical analysis of the properties of the model (1) both at T=0 and for T>0, as a function of $\Delta_o$, particle concentration, interaction parameters and disorder [8]. In the calculations we used for c-electrons a square DOS as well a semielliptic one with the effective bandwidth 2D and assumed the case of alternating (hypercubic) lattices, with $J_{ij}$, $K_{ij}$ and $t_{ij}$ restricted to nearest neighbours ($J_o = \sum_j J_{ij}$, $K_o = \sum_j K_{ij}$).

Below we only quote the main conclusions and show a few representative phase diagrams (Figs 1-4), obtained by a comparison of the free energies $\langle F_o\{E_i\}\rangle_{av}$ of all the possible solutions.

In the absence of the density-density interactions i.e. for V=K=0, the only phases possible in the system are SC and NO, and the effects of disorder in such a case have been discussed in more detail in Ref. [9].

The interactions $K_{ij}$ and $V_o$ can stabilize CDW orderings in the system considered. They compete with SC and in general apart from the SC and NO phases the system can exhibit also the CDW and mixed CDW-SC phases, as well as the phase separated states CDW-SC (PS1) and CDW-NO (PS2).

1) One finds that both the SC and the CDW can be very strongly affected by the diagonal disorder in the local pair subsystem. This is in obvious contrast with the s-wave BCS superconductors which, according to the Anderson's theorem, are rather insensitive to the disorder (i.e. to non-magnetic impurities).

2) The transition between CDW and NO phase a finite temperature can be either second or first order, depending on the strength of disorder. As we see from Fig. 1, for *the two-delta distribution* the increasing $E_o$ changes first the nature of the phase transition from a continuous to discontinuous type, resulting in the tricritical point (TCP), then it suppresses CDW for all T.

3) For $V_o \neq 0$ and/or $K_{ij} > 0$ in a definite range of parameters the system can exhibit the phenomena which we call a *disorder induced superconductivity* and a *disorder induced charge ordering*. In the ground state the former phenomena can occur at half-filling of the both bands ($2n_b = n_c = 1$) if $V_o > I_o$, or $K_o > J_o$: a discontinuous transitions from CDW to SC with increasing $E_o$ (cf. Fig. 4). The latter one can be observed beyond half-filling of local pair band ($2n_b \neq 1$) for the *two-delta distribution* function: the first order transitions from the SC or PS1 states into CDW-NO (PS2) state with increasing $E_o$ (cf. Fig. 3).

4) The detailed features of the phase diagrams are sensitive to the choice of the distribution function of the random potential (see e.g. Figs. 1 and 3). In particular, in the case $J_o \neq 0$, $K_{ij} \neq 0$ ($I_o = V_o = 0$) *for the rectangular distribution* the transitions from SC and CDW phases to NO state at finite temperatures are of second order for any local pair concentration $n_p$ and strength of disorder $E_o$. On the contrary, for *two-delta distribution* both these transitions can be either second or first order, depending on $n_b$ and $E_o$ [8].

5) In the presence of the repulsive density-density interactions there are two types of phase separated states possible in the system considered (PS1 and PS2) which can appear if $2n_b \neq 1$.

In these states, the system breaks into coexisting domains of two different charge densities $n_+$ and $n_-$. In real systems the size of the domains will be finite and determined by the long-range Coulomb repulsion and structural imperfections. We expect, on the basis of our earlier studies concerning the case without disorder, that in the presence of long-range Coulomb interactions the general structure of the phase diagrams derived will remain unchanged except of a possible replacement of PS1 and PS2 states by the homogeneous SC-CDW phase or the incommensurate (or striped) CDW phases.

6) Our results are of relevance for several groups of noconventional superconductors and CDW systems with alternate valence, where the coexistences of LP states with itinerant electron states have been either established or suggested [1-3] and where the disorder introduced by doping or nonstoichiometry has to be an important effect.

**Acknowledgments**

We would like to thank R.Micnas and T.Kostyrko for many useful discussions. This work was supported in part by the Polish State Committee for Scientific Research (KBN), Project No. 2 P03B 154 22.

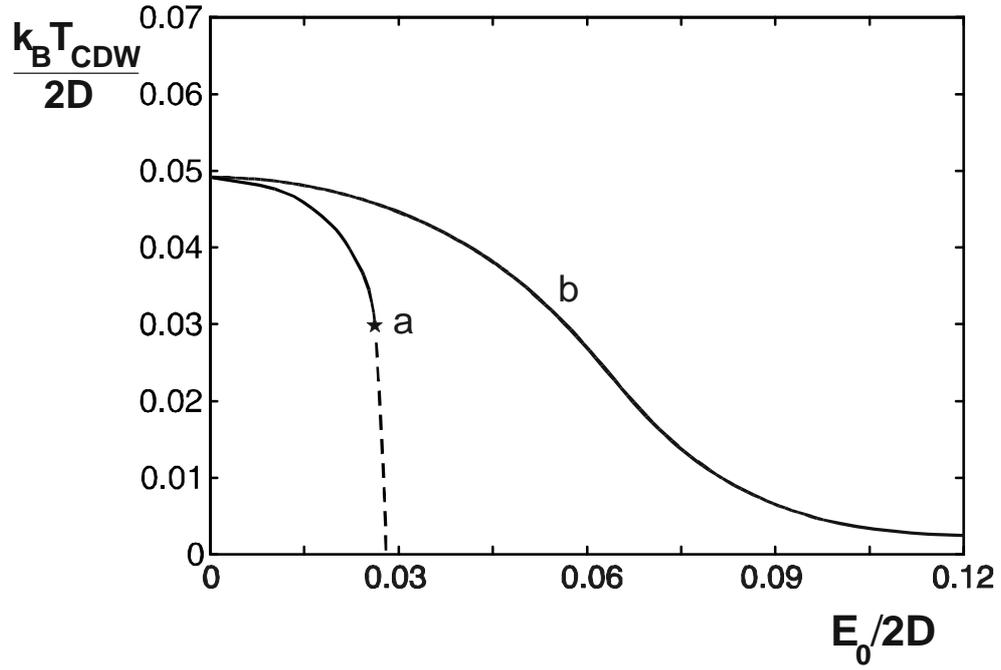

**Fig. 1.** The CDW transition temperatures $k_B T_{CDW}/2D$ as a function of $E_o/2D$ for $n = 2$, $\Delta_o/D = 1$, $V_o/2D = 0.2$, $I_o = J_o = K_o = 0$ plotted for a) *the two-delta distribution*, b) *the rectangular distribution*. Solid and dashed lines mark second- and first order transitions and a star marks a tricritical point (TCP).

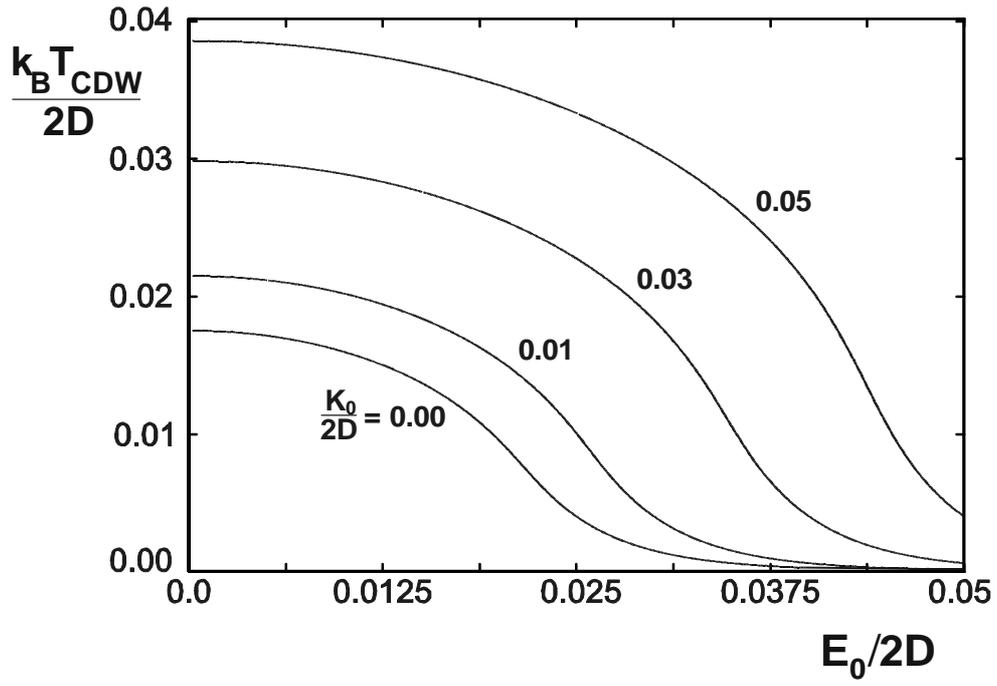

**Fig. 2**. The plots of $k_B T_{CDW}/2D$ vs $E_o/2D$ for $n = 2$, $\Delta_o/D = 1$, $V_o/2D = 0.1$, $I_o = J_o = 0$ and several fixed values of $K_o/2D$ (numbers to the curves). The rectangular distribution of $E_i$.

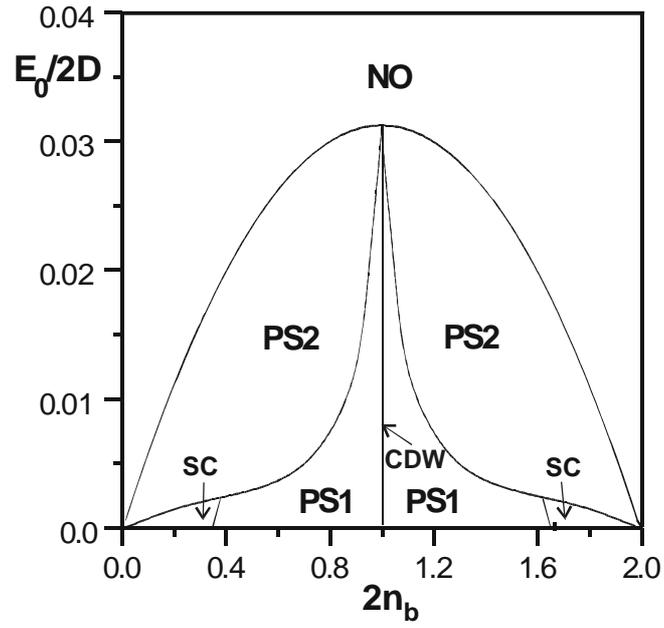

**Fig. 3.** Ground state phase diagrams as a function of $E_o/2D$ and $2n_b$ for $J_o/2D = 0.05$, $I_o = V_o = 0$ and $K_o/J_o = 2.5$ for the two-delta distribution.

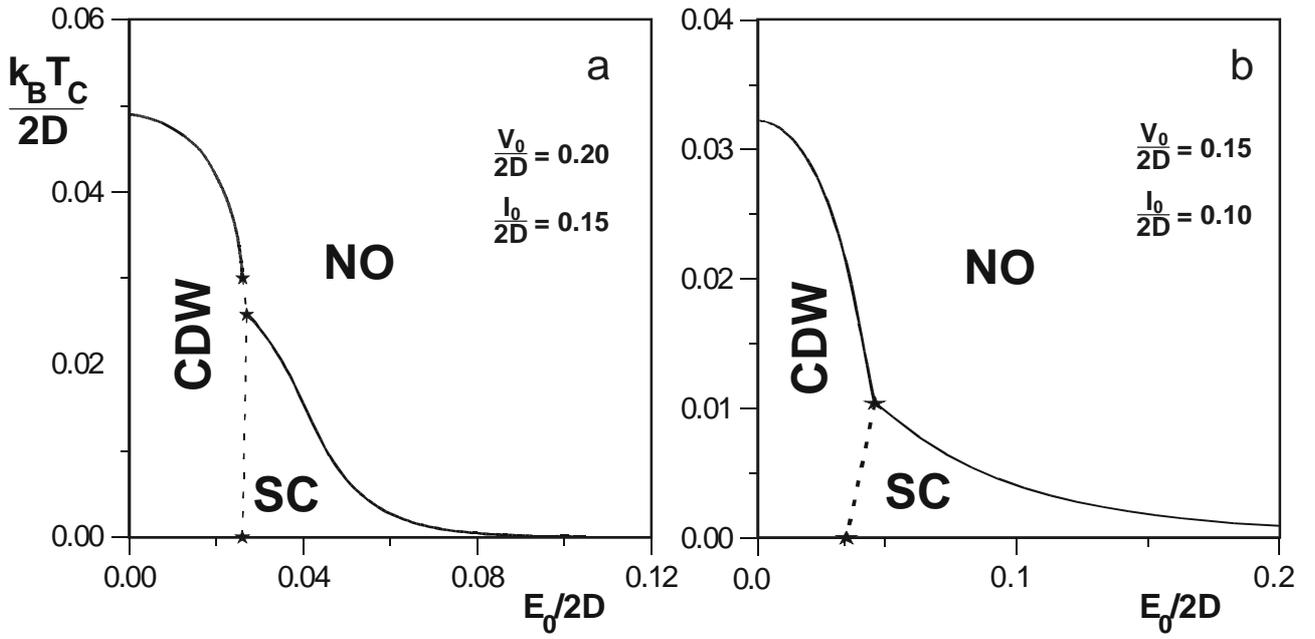

**Fig. 4.** Finite temperature phase diagrams as a function of $E_o/2D$ for $n=2$, $\Delta_o/D=1$, ($J_O = K_O = 0$) plotted for a) the two-delta distribution $I_o/2D = 0.15$ ($V_o/2D = 0.20$), b) the rectangular distribution $I_o/2D = 0.1$ ($V_o/2D = 0.15$).